\documentclass[prl,nofootinbib,floatfix,twocolumn,showpacs,superscriptaddress]{revtex4}
\usepackage[dvips]{pstcol}
\usepackage{graphicx}
\usepackage{dcolumn}
\usepackage{bm}
\usepackage{rotating}
\usepackage{amsmath,amssymb,amsfonts}
\begin{document}

\def\groupargonne{\affiliation{Physics Division, Argonne National Laboratory, Argonne, Illinois 60439-4843, USA}}
\def\groupbari{\affiliation{Istituto Nazionale di Fisica Nucleare, Sezione di Bari, 70124 Bari, Italy}}
\def\groupbeijing{\affiliation{School of Physics, Peking University, Beijing 100871, China}}
\def\groupchina{\affiliation{Department of Modern Physics, University of Science and Technology of China, Hefei, Anhui 230026, China}}
\def\groupcolorado{\affiliation{Nuclear Physics Laboratory, University of Colorado, Boulder, Colorado 80309-0390, USA}}
\def\groupdesy{\affiliation{DESY, 22603 Hamburg, Germany}}
\def\groupzeuthen{\affiliation{DESY, 15738 Zeuthen, Germany}}
\def\groupdubna{\affiliation{Joint Institute for Nuclear Research, 141980 Dubna, Russia}}
\def\grouperlangen{\affiliation{Physikalisches Institut, Universit\"at Erlangen-N\"urnberg, 91058 Erlangen, Germany}}
\def\groupferrara{\affiliation{Istituto Nazionale di Fisica Nucleare, Sezione di Ferrara and Dipartimento di Fisica, Universit\`a di Ferrara, 44100 Ferrara, Italy}}
\def\groupfrascati{\affiliation{Istituto Nazionale di Fisica Nucleare, Laboratori Nazionali di Frascati, 00044 Frascati, Italy}}
\def\groupgent{\affiliation{Department of Subatomic and Radiation Physics, University of Gent, 9000 Gent, Belgium}}
\def\groupgiessen{\affiliation{Physikalisches Institut, Universit\"at Gie{\ss}en, 35392 Gie{\ss}en, Germany}}
\def\groupglasgow{\affiliation{Department of Physics and Astronomy, University of Glasgow, Glasgow G12 8QQ, United Kingdom}}
\def\groupillinois{\affiliation{Department of Physics, University of Illinois, Urbana, Illinois 61801-3080, USA}}
\def\groupmichigan{\affiliation{Randall Laboratory of Physics, University of Michigan, Ann Arbor, Michigan 48109-1040, USA }}
\def\groupmoscow{\affiliation{Lebedev Physical Institute, 117924 Moscow, Russia}}
\def\groupnikhef{\affiliation{Nationaal Instituut voor Kernfysica en Hoge-Energiefysica (NIKHEF), 1009 DB Amsterdam, The Netherlands}}
\def\groupstpetersburg{\affiliation{Petersburg Nuclear Physics Institute, St. Petersburg, Gatchina, 188350 Russia}}
\def\groupprotvino{\affiliation{Institute for High Energy Physics, Protvino, Moscow region, 142281 Russia}}
\def\groupregensburg{\affiliation{Institut f\"ur Theoretische Physik, Universit\"at Regensburg, 93040 Regensburg, Germany}}
\def\grouprome{\affiliation{Istituto Nazionale di Fisica Nucleare, Sezione Roma 1, Gruppo Sanit\`a and Physics Laboratory, Istituto Superiore di Sanit\`a, 00161 Roma, Italy}}
\def\grouptriumf{\affiliation{TRIUMF, Vancouver, British Columbia V6T 2A3, Canada}}
\def\grouptokyo{\affiliation{Department of Physics, Tokyo Institute of Technology, Tokyo 152, Japan}}
\def\groupamsterdam{\affiliation{Department of Physics and Astronomy, Vrije Universiteit, 1081 HV Amsterdam, The Netherlands}}
\def\groupwarsaw{\affiliation{Andrzej Soltan Institute for Nuclear Studies, 00-689 Warsaw, Poland}}
\def\groupyerevan{\affiliation{Yerevan Physics Institute, 375036 Yerevan, Armenia}}
\def\groupnone{\noaffiliation}


\groupargonne
\groupbari
\groupbeijing
\groupchina
\groupcolorado
\groupdesy
\groupzeuthen
\groupdubna
\grouperlangen
\groupferrara
\groupfrascati
\groupgent
\groupgiessen
\groupglasgow
\groupillinois
\groupmichigan
\groupmoscow
\groupnikhef
\groupstpetersburg
\groupprotvino
\groupregensburg
\grouprome
\grouptriumf
\grouptokyo
\groupamsterdam
\groupwarsaw
\groupyerevan


\author{A.~Airapetian}  \groupmichigan
\author{N.~Akopov}  \groupyerevan
\author{Z.~Akopov}  \groupyerevan
\author{M.~Amarian}  \groupzeuthen \groupyerevan
\author{A.~Andrus}  \groupillinois
\author{E.C.~Aschenauer}  \groupzeuthen
\author{W.~Augustyniak}  \groupwarsaw
\author{R.~Avakian}  \groupyerevan
\author{A.~Avetissian}  \groupyerevan
\author{E.~Avetissian}  \groupfrascati
% ...

\collaboration{The HERMES Collaboration} \noaffiliation

\title{Double-hadron leptoproduction in the nuclear medium} 

\date{\today}

\hspace{1cm}

\begin{abstract}
The first measurements of double-hadron production in deep-inelastic scattering 
within the nuclear medium were made with the
HERMES spectrometer at HERA using a 27.6 GeV positron beam. By comparing data
for deuterium, nitrogen, krypton and xenon nuclei,
the influence of the nuclear medium on the ratio of double-hadron
to single-hadron yields was investigated. 
Nuclear effects on the additional hadron are clearly observed, but with
little or no difference among nitrogen, krypton or xenon, and with smaller
magnitude than effects seen on previously measured single-hadron
multiplicities.
The data are compared with models based on partonic energy loss or
pre-hadronic scattering, and with a model based on a purely absorptive 
treatment of the final state interactions. Thus, the double-hadron ratio 
provides an additional tool for studying modifications of hadronization in 
nuclear matter.

\end{abstract}

\pacs{13.87.Fh, 13.60.-r, 14.20.-c, 25.75.Gz}

\maketitle


Hadron production from a free nucleon in deep-inelastic scattering
is generally described by fragmentation functions that contain 
non-perturbative information about parton hadronization. 
These functions are expected to be different for nuclear targets
 \cite{gyulassy} because of several possible effects: energy loss of the 
propagating quarks, rescattering during the pre-hadronic formation process 
or interactions of the final-state hadrons within the nucleus.

Despite recent accurate experimental data from single-hadron leptoproduction 
\cite{her1,her2} and relativistic heavy-ion collisions 
\cite{star,phenix,phobos}, the underlying mechanisms
in theoretical models for hadronization in the nuclear medium
differ greatly \cite{xnw,arleo,k1,falter,accardi}. 
In-medium modification of the quark fragmentation function has been described 
in terms of rescattering of gluons and quarks, and of energy loss due to 
induced gluon radiation \cite{xnw,arleo}.
Alternatively, colorless pre-hadron rescattering in the medium has been
suggested \cite{k1,falter} with additional effects due to $Q^2$ rescaling 
\cite{accardi}. Older interpretations \cite{gyulassy,bial} based on hadronic 
final-state interactions require a hadron formation length smaller than the 
nuclear size, which is unlikely for struck quarks boosted to energies larger 
than a few GeV. Although models based on some of these ideas are already in
conflict with data, clearly other types of data are needed to further distinguish
among these interpretations.

Double-hadron leptoproduction offers an additional way to study hadronization.
If partonic energy loss of the struck quark were the only mechanism
involved, it would be na\"{i}vely expected that the attenuation effect does
not depend strongly on the number
of hadrons involved, and the double-hadron to single-hadron ratio for a
nuclear target should be only slightly dependent on the mass number $A$.
On the contrary, if final hadron absorption were the dominant process, 
the requirement of an additional slower sub-leading hadron that is more 
strongly absorbed would suppress the two-hadron yield from heavier nuclei 
\cite{k1,w1}, so that this ratio should decrease with $A$.

Data from STAR \cite{st1} on hadron pair production as a function of azimuthal 
angle showed that for a fixed value of the trigger hadron's transverse 
momentum, the production of opposite-side hadron pairs is completely suppressed for 
central Au+Au collisions due to the final-state interactions with the dense 
medium  generated in such collisions. On the other hand, the same-side pairs 
exhibit jet-like correlations that are similar to p+p and p+d collisions.
These results were used in Ref.~\cite{xnwb} to advocate the picture
that jet fragmentation occurs outside the dense medium. In this model
it has been shown that if hadron absorption or rescattering were
responsible for the observed hadron suppression, it would likely destroy the 
jet structure, and in particular the correlations between leading and 
sub-leading hadrons within the jet cone. However, the heavy-ion data cannot 
exclude hadronic absorption effects completely. 

In this paper the first measurement of double-hadron leptoproduction
on nitrogen, krypton and xenon relative to deuterium is presented. All charged 
hadrons and $\pi^{0}$ mesons are considered.

Semi-inclusive deep-inelastic scattering data
are presented in terms of the ratio

\begin{equation}
R_{2h}(z_2)=\frac{\large(\frac{dN^{z_1>0.5}(z_2)/dz_2}{N^{z_1>0.5}}\large)_A}
{\large(\frac{dN^{z_1>0.5}(z_2)/dz_2}{N^{z_1>0.5}}\large)_D},
\label{r2h}
\end{equation}

\noindent in which $z=E_h/\nu$ is the fractional hadronic energy,  $E_h$ is the
hadron energy and $\nu$ is the virtual photon energy, all of which are 
evaluated in the target rest frame.
The values $z_1$ and $z_2$ correspond to the leading (largest $z$) and 
sub-leading (second largest $z$) hadrons, respectively.
The quantity $dN^{z_1>0.5}$ is the number of events with at least two
detected hadrons in a bin of width $dz_2$ at  $z_2$ with $z_1>0.5$.  The 
quantity  $N^{z_1>0.5}$ is the number of events with at least one detected 
hadron with $z_1>0.5$.  The label $A(D)$ indicates that the term is calculated 
for a nuclear (deuterium) target.

The measurement was performed with the HERMES 
spectrometer~\cite{hermesdetector} using the 27.6 GeV positron beam stored in 
the HERA ring at DESY. The spectrometer consists of two identical halves 
located above and below the positron beam pipe. The scattered positrons and the
produced hadrons were detected simultaneously within an angular acceptance  of 
$\pm$ 170 mrad horizontally, and $\pm$ (40 -- 140) mrad vertically.

The nuclear targets, which were internal to the positron
storage ring, consisted of polarized or unpolarized deuterium, or unpolarized 
high-density nitrogen, krypton or xenon gas injected into a 40 cm long open-ended
tubular storage cell. Target areal densities up to  1.4 $\times$ 10$^{16}$ 
nucleons/cm$^2$ were obtained for unpolarized gas corresponding to 
luminosities up to 3 $\times$ 10$^{33}$ nucleons/(cm$^2$ s). 

The positron trigger was formed by a coincidence between the signals from three 
scintillator hodoscope planes, and a lead glass calorimeter where a minimum 
energy deposit of 3.5 GeV (1.4 GeV) for unpolarized (polarized) 
target runs was required. The scattered positrons were identified using
a transition-radiation detector, a scintillator pre-shower counter, and an 
electromagnetic calorimeter.
Scattered positrons were selected by imposing constraints on the squared 
four-momentum of the virtual photon $Q^2>1$~GeV$^2$, on the invariant
mass of the photon-nucleon system $W=\sqrt{2M\nu+M^2-Q^2} >$2 GeV
where $M$ is the nucleon mass, and on the energy fraction of the virtual 
photon $y=\nu/E<$~0.85 where $E$ is the beam energy.
The constraints on $W$ and $y$ are applied to exclude nucleon resonance
excitations and to limit the magnitude of the radiative corrections, 
respectively.
In addition the requirement $\nu>$7 GeV was imposed to limit the 
kinematical correlations between $\nu$ and $z$.

Charged hadrons (i.e. $\pi$, $K$ and $p$ without distinction) were 
reconstructed for momenta above 1.4 GeV.
The electromagnetic calorimeter~\cite{calo} provided neutral pion 
identification through the detection of neutral clusters originating from 
two-photon decay.
Each of the two clusters was required to have an energy $E_{\gamma}\ge 0.8$ GeV.
The $\pi^0$ mesons were selected by requiring that the reconstructed
invariant mass was within two standard deviations of the center of the
$\pi^0$ mass peak.

The leading hadron was selected with $z_1>$0.5. In this case, 
it is expected to contain the struck current quark with high 
probability. No explicit constraint was applied to $z_2$.
Both $z_1$ and $z_2$ were calculated assuming that all hadrons
have the mass of the pion.

Using the code of Ref.~\cite{Tera}, radiative corrections to $R_{2h}$ were 
found to be negligible in the whole kinematic range. This is because there is
no elastic or quasi-elastic tail in semi-inclusive events, and the inelastic 
corrections largely cancel in the measured ratio.

Two methods of double-hadron event selection were used.
Selection I contains only the combinations of hadron charges 
(leading-subleading) $++$, $--$, $+0$, $0+$, $-0$, $0-$, $00$.
This suppresses the contributions from $\rho^0 \rightarrow \pi^+ \pi^-$ decay
because the $+-$ and $-+$ combinations are missing. 
Moreover, in the Lund string model, the exclusion of the opposite-charge
combinations enhances the rank-1 (leading) plus rank-3 (sub-leading) 
combination \cite{lund}. The higher the particle rank,
the more likely it is formed deep inside the nucleus, and the corresponding
hadron absorption should be larger.
Selection II contains all particle charge combinations.
Here, the sub-leading hadron is mainly of rank-2 and the contribution from
$\rho^0$ decay is larger. In both Selections I and
II the relative yield from exclusive $\rho^0$ production in $N^{z_1>0.5}$
is small and was evaluated by Monte Carlo calculation to be on the order of 12\% 
for the deuterium target.

\begin{figure}[ht]
\includegraphics[height=4.5in,width=4.5in]{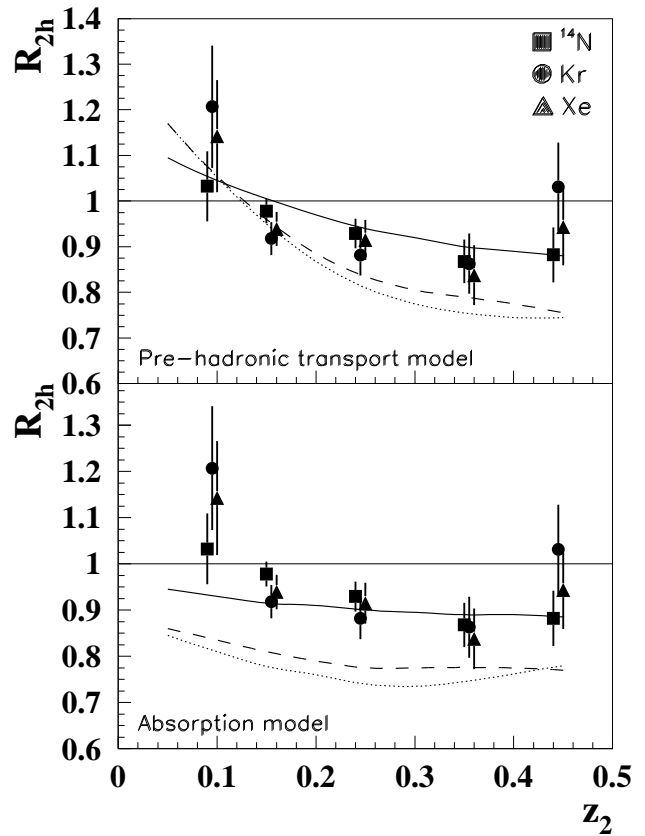}
\caption{\label{f1}The ratio $R_{2h}$ as a function of $z_2$ for $^{14}$N 
(squares), Kr (circles) and Xe (triangles) with $z_1>$0.5. Only Selection I 
was considered. The systematical uncertainty is 2\% for all the targets 
and is independent of $z_2$. In the upper panel the curves (solid 
for $^{14}$N, dashed for Kr, dotted for Xe) are calculated within a BUU 
transport model \cite{falter}. In the bottom panel the same data are shown with 
calculations that assume only absorption for the three nuclei (same line types 
as in the upper plot) \cite{falter}.}
\end{figure}

Fig.~\ref{f1} shows the double ratio $R_{2h}$ as a function
of $z_2$ for Selection I only.
The kinematic variables are in the range $\langle \nu \rangle$=21 to 16 GeV and
$\langle Q^2 \rangle$=2.1 to 2.6 GeV$^2$ as $z_2$ goes from 0.09 to 0.44.
The averages over $z_2$ are $\langle \nu \rangle=$17.7 GeV and 
$\langle Q^2 \rangle=$2.4 GeV$^2$.

\begin{figure}[ht]
\includegraphics[height=2.85in,width=4.5in]{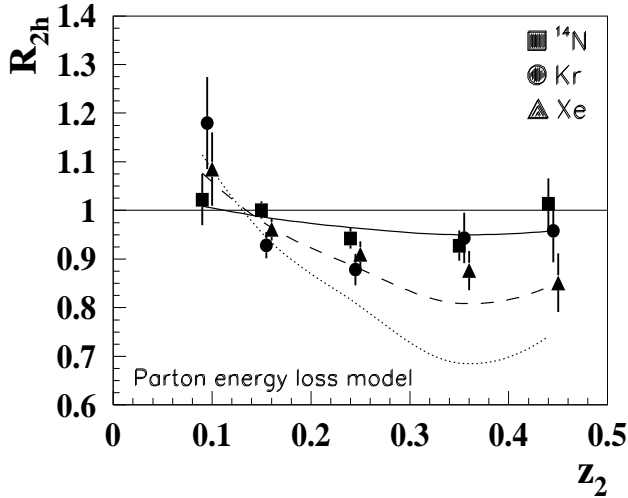}
\caption{\label{f2}The ratio $R_{2h}$ as a function of $z_2$ for $^{14}$N 
(squares), Kr (circles) and Xe (triangles) with $z_1>$0.5 for Selection II.
The systematic uncertainty is 4\% (3\%) for xenon and krypton (nitrogen) and is
independent of $z_2$. The curves ($^{14}$N: solid; Kr: dashed; Xe: dotted) are 
from the parton energy loss model described in Refs.\cite{xnwb,am}.}
\end{figure}

The ratio $R_{2h}$ is generally below unity with no significant difference 
between the three nuclei.
These data clearly show that the nuclear effect in the double-hadron ratio
is much smaller than for the single-hadron attenuation measured under the
same kinematic conditions \cite{her1,her2}.
For $z_2<$0.1, where $R_{2h}$ rises towards and possibly above 1, the slow 
hadrons originate largely from target fragmentation \cite{falter,tfth}. Also 
for $z_2>$0.4, where the two hadrons have similar energy, $R_{2h}$ seems 
to rise towards unity.

Fig.~\ref{f1} (upper panel) shows calculations based on a PYTHIA event generator
with a fully coupled-channel treatment of final-state interactions by means
of a BUU transport model \cite{falter}. In this model, the fragmentation 
function is modified by pre-hadron interactions and rescattering in the medium.
Although the general trend of the data is reproduced, the model predicts an 
effect twice as large for xenon and krypton as for nitrogen above $z_2$=0.1, 
which is not supported by the data.

Fig.~\ref{f1} (bottom panel) shows the same data compared to a calculation
with a purely absorptive treatment of the interaction of the pre-hadronic or 
the final hadronic states. The data rule out this assumption \cite{falter}. 

Fig.~\ref{f2} presents $R_{2h}$ calculated for all hadron charge combinations 
(Selection II). Inclusion of the $+-$ and $-+$ pairs does not change
the value of $R_{2h}$ significantly, contrary to all of the na\"{i}ve space-time
evolutionary models of hadronization. In order to evaluate further the possible 
influence of the exclusive and semi-inclusive $\rho^0$ production, $R_{2h}$ 
was extracted for all hadron pairs 
except those with invariant mass near the $\rho^0$. This has no noticeable 
effect on $R_{2h}$. Therefore, the final data are presented over the full 
invariant mass range. The effect of only the exclusive $\rho^0$ production
is even smaller since it contributes only 5\% of the total yield. 
The contamination from exclusive production of $\rho^{\pm}$ and $\omega$
mesons is completely negligible, being suppressed by more than
one order of magnitude with respect to the $\rho^0$ \cite{md}.

The curves in Fig.~\ref{f2} represent the model \cite{xnwb,am} in which 
modifications of the 
fragmentation functions arises from parton energy loss.  
Contrary to na\"{i}ve expectations, this model predicts a significant 
$A$-dependence, in conflict with the data.

Table~\ref{integr} provides a quantitative comparison between the data and 
theoretical predictions for $R_{2h}$ integrated over $z_2$.

The total systematic uncertainty on $R_{2h}$ is 4\% (3\%) 
for xenon and krypton (nitrogen) and is nearly independent of $z_2$.
The main contribution to the systematic uncertainty comes from the decay of
exclusively produced $\rho^0$ mesons. However, 
for the double-hadron multiplicities $dN^{z_1>0.5}(z_2)/dz_2$
the $\rho^0$ contribution has a negligible effect. The $\rho^0$ contribution to 
$N^{z_1>0.5}$ was estimated in analogy with Ref.\cite{her1,her2} to be
about 2\% (3\%) for light (heavy) nuclei. 
The only other contributing factor is the uncertainty in the overall efficiency
of 2\%.
The geometric acceptance for semi-inclusive hadron production was verified
to be the same for both the nuclear and deuterium targets by studying the 
multiplicity ratio as a function of the hadron polar angle.
This ratio is constant within experimental error.

In conclusion, the first measurement of double-hadron production on deuterium, 
nitrogen, krypton and xenon is presented. The ratio of double- to 
single-hadron yields from nuclear targets compared to deuterium are 
similar for atomic mass numbers $A$=14, 84 and 131, as a function of the 
relative energy of the sub-leading hadron.  
This is at variance with the single-hadron attenuation data, which depend
strongly on $A$.  The data do not support na\"{i}ve expectations for 
pre-hadronic and hadronic final-state interactions that are purely absorptive.
Models that interpret modifications to fragmentation as being due to
pre-hadronic scattering or partonic energy loss are also inconsistent with
the data. In fact the latter predict an even larger $A$-dependence, while
the data show little. Like the jet correlation
measurements in heavy-ion collisions, the double-hadron observables
in semi-inclusive deep inelastic scattering provide new information for 
differentiating between models of hadronization in nuclei that are 
indistinguishable in single-hadron measurements.

\vspace{1.0cm}

We are grateful to T.Falter, B.Kopeliovich, A.Majumder and X.N.Wang
for useful discussions.
We gratefully acknowledge the DESY management for its support, the staff
at DESY and the collaborating institutions for their significant effort,
and our national funding agencies and the EU RII3-CT-2004-506078 program
for financial support.

\begin{widetext}
\begin{center}
\begin{table}[h]
\caption{\label{integr}Averaged values of $R_{2h}$ for $z_1>$0.5.}
\begin{tabular}{|c|c|c|c|c|c|}
\hline \hline
& $z_2<$0.5 & 0.1$<z_2<$0.5 & 0.1$<z_2<$0.5 & 0.1$<z_2<$0.5 & 0.1$<z_2<$0.5  \\ 
& & &  Model \cite{falter} &  Model Abs. \cite{falter} & Model \cite{xnwb,am} \\ \hline
$^{14}$N/D & & & & & \\
~Selection I~ & 0.946 $\pm$ 0.017 $\pm$ 0.019 & 0.941 $\pm$ 0.018 $\pm$ 0.019 & 0.931 & 0.907 & - \\ \hline
$^{14}$N/D & & & & & \\
~Selection II~ & 0.975 $\pm$ 0.009 $\pm$ 0.029 & 0.972 $\pm$ 0.010 $\pm$ 0.029 & - & - & 0.965 \\ \hline
Kr/D & & & & & \\
~Selection I~ & 0.929 $\pm$ 0.015 $\pm$ 0.019 & 0.917 $\pm$ 0.016 $\pm$ 0.018 & 0.835 & 0.796 & - \\ \hline
Kr/D & & & & & \\
~Selection II~ & 0.902 $\pm$ 0.008 $\pm$ 0.036 & 0.892 $\pm$ 0.008 $\pm$ 0.036 & - & - & 0.879 \\ \hline
Xe/D & & & & & \\
~Selection I~ & 0.936 $\pm$ 0.023 $\pm$ 0.019 & 0.915 $\pm$ 0.024 $\pm$ 0.018 & 0.815 & 0.773 & - \\ \hline
Xe/D & & & & & \\
~Selection II~ & 0.936 $\pm$ 0.012 $\pm$ 0.037 & 0.925 $\pm$ 0.013 $\pm$ 0.037 & - & - & 0.800 \\ \hline \hline
\end{tabular}
\end{table}
\end{center}
\end{widetext}

\noindent (*)Permanent address: Thomas Jefferson National Accelerator Facility, 
Newport News, Virginia 23606, USA.
(**)Permanent address: College of William $\&$ Mary, Williamsburg, Virginia 
23187, USA.

\end{document}